\documentclass[groupedaddress,
 reprint,
 amsmath,amssymb,
 aps,
 prd,
 nofootinbib
]{revtex4-1}

\usepackage{graphicx}
\usepackage{dcolumn}
\usepackage{bm}
\usepackage{hyperref}
\usepackage[utf8]{inputenc}
\usepackage[capitalise]{cleveref}
\usepackage{xcolor}

\usepackage{aas_macros}

\begin{document}

\title{Formation and structure of ultralight bosonic dark matter halos}

\author{Jan Veltmaat}
\email{jvelt@astro.physik.uni-goettingen.de}
\author{Jens C. Niemeyer}
\email{jens.niemeyer@phys.uni-goettingen.de}
\author{Bodo Schwabe}
\email{bschwabe@astro.physik.uni-goettingen.de}
\affiliation{%
 Institut f\"ur Astrophysik\\
 Universit\"at G\"ottingen
}%

\date{\today}

\begin{abstract}
We simulate the formation and evolution of ultralight bosonic dark matter halos from cosmological initial conditions. Using zoom-in techniques we are able to resolve the detailed interior structure of the halos. We observe the formation of solitonic cores and confirm the core-halo mass relation previously found by Schive et al. The cores exhibit strong quasi-normal oscillations that remain largely undamped on evolutionary timescales. On the other hand, no conclusive growth of the core mass by condensation or relaxation can be detected. In the incoherent halo surrounding the cores, the scalar field density profiles and velocity distributions show no significant deviation from collisionless N-body simulations on scales larger than the coherence length. Our results are consistent with the core properties being determined mainly by the coherence length at the time of virialization, whereas the Schrödinger-Vlasov correspondence explains the halo properties when averaged on scales greater than the coherence length.
\end{abstract}

\maketitle

\section{Introduction}
\label{sec:intro} 

The hypothesis that dark matter is composed of an ultralight bosonic field with particle mass $m \gtrsim 10^{-22}$ eV is well-motivated from the point of view of fundamental theories with weakly broken shift symmetries, where they naturally occur as axion-like particles \cite{Arvanitaki2010,Marsh2015a,Hui2017}. In these scenarios, their self-interaction can typically be neglected for questions of structure formation, making them candidates for ultralight scalar, or ``fuzzy'', dark matter (FDM) \cite{Sin1994,Hu2000}. 

In spite of the simplicity of FDM which is fully specified by the single parameter $m$, its fundamentally wave-like behavior on scales near the de Broglie wavelength gives rise to interesting new phenomena that can potentially be probed by purely gravitational interactions. In their pioneering work, Schive \textit{et al.} \cite{Schive2014} used adaptively refined simulations of the Schrödinger-Poisson equations to resolve the density structure of collapsed dark matter halos for the first time. Their results revealed the formation of coherent, solitonic cores embedded in incoherent, granular halos with CDM-like density profiles. The inner dark matter profile of dwarf galaxies can be probed by stellar kinematics and so far observations favour a cored central density profile over the cuspy profile predicted by CDM simulations \cite{Gilmore2007, Blok2010}. FDM thus potentially solves the cusp-core problem of the $\Lambda$CDM model if $m \sim 10^{-22}$ eV \cite{Hu2000,Peebles2000,Marsh2015c,Calabrese2016,Chen2016,Gonzales-Morales:2016mkl} which is, however, in tension with the reported lower bound from the Lyman-alpha forest of $m > 2\times10^{-21}$eV \cite{Irsic2017,Kobayashi2017} and may not explain the observed core density profiles \cite{Deng2018}. Constraints from the cosmic microwave background and large scale structure \cite{Hlozek2015,Hlozek2017,Hlozek2018,Nguyen2017}, high-$z$ galaxy luminosity functions and  reionization \cite{Bozek2015,Schive2016,Corasaniti2017}, or the abundance of dark matter halos and subhalos \cite{Marsh2014,Du2017,Menci2017,Schive2018} are similar but weaker by about an order of magnitude. A recent investigation of the stability of FDM subhalos in the Milky Way suggests a bound closer to the Lyman-alpha forest result \cite{Du2018}.
Several different scenarios for the formation of FDM cores and the dynamics governing their time evolution have been proposed.
Schive \textit{et al.} \cite{Schive2014a} found that the core masses obey a scaling relation with the mass of their host halos that was explained heuristically with a variant of the uncertainty relation.
In \cite{Hui2017}, core formation is explained as a relaxation process of quasi-particles with sizes given by their de Broglie wavelength. Following this idea, \cite{vicens2018} predicted the core mass of a halo depending on its age. Wave condensation described by kinetic theory was considered in \cite{Levkov2018} as the relevant mechanism for core formation. Using the empirical equivalence between averaged radial profiles of FDM and N-body halos on scales greater than the coherence length, \cite{Mocz2018} derive a relation between core and halo mass and argue that it holds near the classical limit. Virialized halos of Milky Way-sized galaxies have also been investigated using self-consistent constructions with prescribed distribution functions \cite{Lin2018}.

As for CDM, detailed insight into structure formation with FDM relies heavily on numerical simulations. Depending on the particular goals, different numerical techniques have been employed. Arguably, the most precise method is to directly solve the Schr\"odinger-Poisson (SP) equations using either finite-difference \cite{Schive2014,Schwabe2016} or pseudo-spectral methods \cite{Woo2009,Mocz2017,Mocz2018,Du2018}. Explicitly resolving the granular interference structure on scales of the spatial and temporal coherence length, these methods require large computational resources. On the other hand, when focusing on the impact of the linear suppression of small-scale power in FDM cosmologies, the correspondence of the coarse-grained SP and Vlasov-Poisson (VP) equations \cite{Widrow1993,Uhlemann2014} on large scales permits the use of standard N-body simulations with FDM initial conditions \cite{Schive2016,Sarkar2016,Irsic2017,Kobayashi2017}. An intermediate approach is to modify N-body or SPH codes with an additional force term in order to capture the effects of the scalar field gradient energy \cite{Mocz2015,Veltmaat2016,Baldi2018}. While these methods have lower demands on resolution than direct SP simulations, they struggle to accurately treat interference phenomena of FDM.

The focus of this work lies on the cosmological evolution of FDM halos, including the formation and time dependence of solitonic cores and the statistical distribution of density and velocity in the incoherent halo outside of the core. Both phenomena are intrinsically wave-like effects that depend on the halo's formation history. Our simulations therefore require both a sufficiently large spatial domain with cosmological initial conditions to simulate realistic halos, and an accurate representation of the scalar field by solving the SP equations inside the halo. Regions outside of collapsed regions, on the other hand, to reasonable approximation only affect the large-scale gravitational field and the mass accretion onto the halo. Providing appropriate boundary conditions we can thus treat them with N-body dynamics, saving considerable computational costs. Adaptive-mesh refinement (AMR) offers a suitable framework for combining both approaches in a hybrid SP/N-body scheme.

The remainder of this paper is structured as follows. After describing our numerical method in \cref{sec:methods}, we report the results of our simulations in \cref{sec:results}, followed by a discussion of core formation, time dependence, and halo properties in the context of known theoretical results. We conclude in \cref{sec:conclusions}.

\section{Methods}
\label{sec:methods}

The time evolution of the nonrelativistic FDM density $\rho = |\Psi|^2$ obeys the comoving SP equations,
\begin{align}
\label{eq:schroedinger}
 i\hbar \frac{\partial \Psi}{\partial t} = -\frac{\hbar^2}{2ma^2} \nabla^2 \Psi + V m \Psi
\end{align}
and
\begin{align}
 \nabla^2 V = \frac{4\pi G}{a} \left(\rho-\overline{\rho}\right)\,\,,
\end{align}
with the scale factor $a$ and the mean cosmic density $\bar \rho$.

Our simulations are performed with a modified version of the public cosmology code \textsc{Enzo}\footnote{http://enzo-project.org} \cite{Bryan2014}. Most of the simulation domain is simulated by the standard N-body procedure already implemented in \textsc{Enzo} to simulate CDM. In each simulation, one individual halo is resolved further by additional refinement levels (see below for further details). On the most refined level, called the ``Schrödinger domain'' in the following, we solve the SP equations using the 4th-order Runge-Kutta solver that was employed in \cite{Schwabe2016}. A crucial point in this approach is the treatment of the boundary conditions of the Schr\"odinger domain where infalling particles are converted into a representation of the wavefunction $\Psi$.

We use the ``classical wave function" formulation \cite{Trahan2005} for the initial conditions and boundaries of the Schr\"odinger domain. The classical wave function approximates the actual Schr\"odinger wave function under the assumption that interference effects are negligible. In this representation, particles carry a classical phase $S_{i}$ which is evolved according to the Hamilton-Jacobi equation\cite{Trahan2005}
\begin{align}
\label{eq:hamilton-jacobi}
    \frac{\text d S_{i}}{\text d t} = \frac{1}{2} \mathbf{v_i}^2 - V(\mathbf{x_i})~,
\end{align}
where $\mathbf{v_i}$ and $\mathbf{x_i}$ are the velocity and location of the $i$th particle, respectively. 

Before each Runge-Kutta time step on the most refined level, the classical wave function
\begin{align}
\label{eq:classwfunc}
    \Psi_\mathrm{c}(\mathbf{x}) = R_\mathrm{c}(\mathbf{x})~ e^{i S_\mathrm{c}(\mathbf{x})m/\hbar}~,
\end{align}
is constructed at the boundaries. Using a second order interpolation kernel,
\begin{align}
 W(|\mathbf{x} - \mathbf{x_i}|) = m_i \frac{3}{\pi \xi^3} \left(1-\frac{|\mathbf{x} - \mathbf{x_i}|}{\xi}\right)
\end{align}
for $|\mathbf{x} - \mathbf{x_i}| < \xi$ and $0$ elsewhere, the classical amplitude is given by 
\begin{align}
\label{eq:rcl}
   R_\mathrm{c}(\mathbf{x}) = \sqrt{\sum_i W(\mathbf{x} - \mathbf{x_i})}\,\,.
\end{align}
$S_\mathrm{c}$ is interpolated from the particle positions to the grid by taking the complex phase of the dummy field  
\begin{align}
\label{eq:dummy}
   \Psi_\mathrm{d}(\mathbf{x}) &= \sum_i \sqrt{W(\mathbf{x} - \mathbf{x_i})} e^{i(S_i + \mathbf{v_i}\cdot a(\mathbf{x}-\mathbf{x_i}))m/\hbar}\nonumber\\ 
   &= R_\mathrm{d}(\mathbf{x})~ e^{i S_\mathrm{c}(\mathbf{x})m/\hbar} \,\,.
\end{align}
Note that \cref{eq:hamilton-jacobi} describes the evolution of the central phase of a localized wave package moving in a potential $V$ according to \cref{eq:schroedinger}. \Cref{eq:dummy} can therefore be understood as the superposition of particles acting like wave packages. By using $R_\mathrm{c}$ instead of $R_\mathrm{d}$ in \cref{eq:classwfunc}, we erase interference patterns in the superposition but ensure mass conservation. Although the classical wave function is only similar to the exact solution of \cref{eq:schroedinger}, if interference is not present (in the single stream regime), it can still provide appropriate boundary conditions for the Schr\"odinger domain when the condition is not necessarily fulfilled, i.e. at late cosmological times. Most importantly, the method yields the correct rate of mass inflow, as we checked by comparing the total mass of the Schr\"odinger field with the total particle mass in the Schr\"odinger domain, showing a maximum deviation of 10\% at all times in all simulations. Thus, the simulations in the Schr\"odinger domain reproduce the generic and statistical properties of FDM halos with the given accretion history. Because of the significant approximations in the boundary conditions however, we do not claim that fine-grained properties like positions of individual granules at any given time are the exact solution for the given initial conditions.

The smoothing radius $\xi$ must be chosen to provide a sufficiently smooth interpolation of the particle density. We used $\xi = 8 \Delta x$ where $\Delta x$ is the cell width at the most refined level. We checked that increasing the radius further does not systematically lead to different results. However, the core mass in \cref{fig:plot4} can differ by up to 30\% owing to the approximations in the employed boundary conditions.

Particles inside the Schr\"odinger domain are evolved further but do not contribute to the density field that sources gravity. Instead, the density of the Schr\"odinger field $|\Psi|^2$ acts as a source of gravity in this region. 

\subsection{Simulation Setup}

We generate initial conditions with \textsc{Music} \cite{Hahn2011} using a transfer function for FDM generated by AxionCAMB \cite{Hlozek2017}. All our simulations have a side length of 2.5 Mpc/h. We choose $H_0 = 70$ km/s/Mpc, $\Omega_\Lambda = 0.75$, $\Omega_m = \Omega_{FDM} = 0.25$ and $m_{22} = m/(10^{-22}$ eV) $ = 2.5$. Starting from redshift $z = 60$ we sample phase space with $\sim 2.8\times 10^{8}$ particles.

Employing the Poisson solver implemented in \textsc{Enzo}, the initial particle phases $S_{i}$ are computed by solving
\begin{align}
    \nabla\cdot\mathbf v = a^{-1}\nabla^{2} S
\end{align}
and interpolating from the grid to the particle positions. Here, $\mathbf v$ is the velocity field generated by \textsc{Music}.

On top of the root grid with $512^3$ cells, two nested static refinement levels with a side length of roughly a quarter of the total domain are centered on the Lagrangian patch of a previously chosen halo. Three additional refinement levels with side lengths of 0.0625 Mpc/h trace the position of the halo's maximum density. Using a refinement factor of two between levels, we resolve the finest one with  a cell width of 150 pc/h. In order to determine the halo's Lagrangian patch and the position of its maximum density over time, we run low resolution standard N-body simulations.

To minimize computational cost, the SP solver is applied only after a redshift of $z \approx 7$, where the particles are still in the single stream regime
and the gradient energy of $\Psi$ is negligible. At this redshift, the classical wave function is constructed at the most refined level and serves as an initial condition for the SP solver.
Like for the smoothing radii, initializing at earlier times has no systematic effects but produces statistical scattering of the resulting core mass of 30\%.

In total we have simulated seven halos with a mass range between $8 \times 10^8$ M$_\odot$ and $7 \times 10^{10}$ M$_\odot$. For comparisons with standard CDM dynamics, we have rerun five of these simulations with only the N-body solver using identical grid resolution and level setup.

\section{Results}
\label{sec:results}

\begin{figure}
\includegraphics[width=\columnwidth]{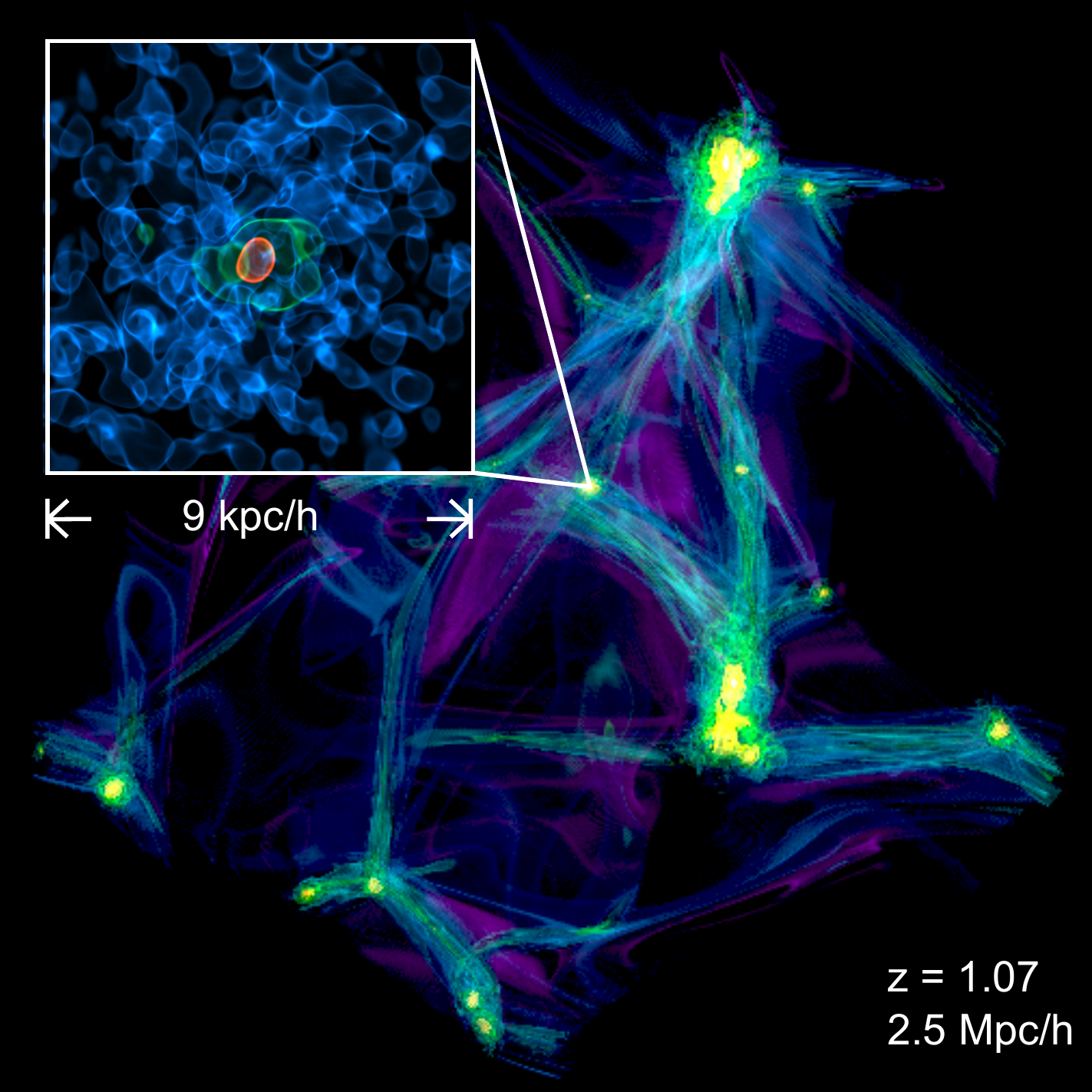}
\caption{Volume rendering of a typical simulation. The large box shows the N-body density in the full simulation domain, the inlay shows the density of the Schr\"odinger field in the central region of the indicated halo. The density thresholds in the inlay are set to 0.75, 0.05 and 0.01 times the maximum density.} 
\label{fig:plot1}
\end{figure}

For this work, we only consider halos that evolve without major mergers. These are more abundant in FDM cosmologies relative to CDM, owing to the low-mass cut-off in the initial power spectrum. \Cref{fig:plot1} shows a typical snapshot of our simulations.

\subsection{Averaged properties}

\begin{figure}
\includegraphics[width=\columnwidth]{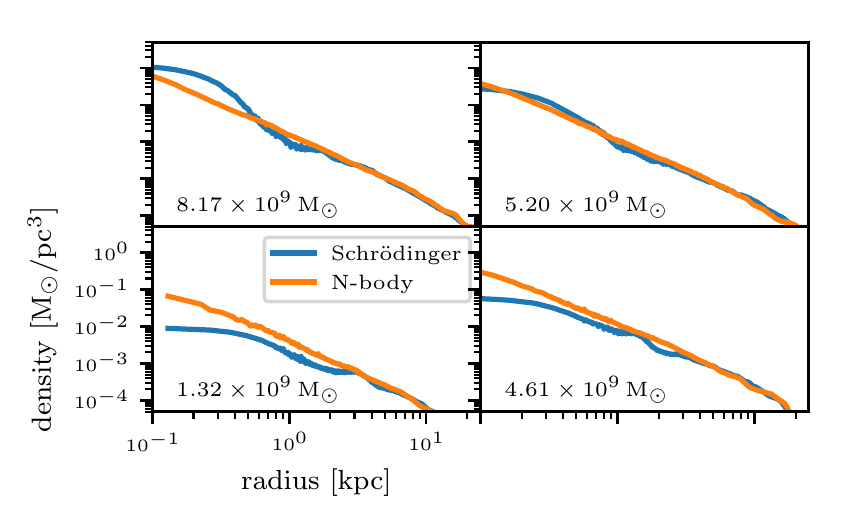}
\caption{Radial FDM and CDM density profiles of four representative halos. The labels indicate their virial masses.
}
\label{fig:plot2a}
\end{figure}

Radial density profiles centered around the maximum density of four representative halos are compared with results from pure N-body runs in \cref{fig:plot2a}. Here, the virial mass of a halo is the mass enclosed by the virial radius, $r_\mathrm{vir}$,  defined as the radius where the enclosed mean density is equal to $\zeta(a) \bar \rho$ with \cite{Bryan1998}
\begin{align}
\label{eq:zeta}
    \zeta(a) \Omega_m(a) = 18\pi^2 + 82 \left(\Omega_m(a)-1 \right)- 39\left(\Omega_m(a)-1\right)^2\,.
\end{align}
Taking radial density profiles already involves smoothing the density by averaging over spherical shells. Consequently, the granular structure of FDM halos which deviates strongly from the smooth CDM density field on small scales, is not visible apart from a small region around the solitonic core. The radially averaged core profile agrees well with previous results \cite{Schive2014,Schwabe2016,Mocz2017}. Among the five halos in our sample that were rerun with a pure N-body solver, the maximum FDM core density was exceeded by the maximum density of the corresponding CDM halo in all cases but one, by a maximum factor of 7.5. The density at the outer edge of the core is not a constant fraction of the maximum core density in our simulations. Their ratio depends both on the phase of the large-amplitude core oscillations discussed below and the time since core formation, with a trend towards decreasing values. It varies between initial values of roughly 0.05 and 0.15 and reaches a minimum value of 0.015 in one of the simulated halos. 

\begin{figure}
\includegraphics[width=\columnwidth]{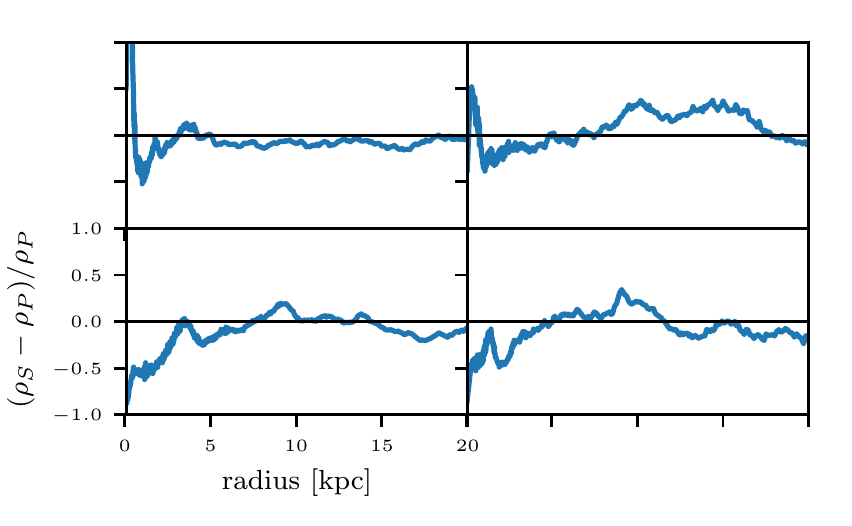}
\caption{Residuals of the same profiles as in \cref{fig:plot2a} in a linear plot.}
\label{fig:plot2b}
\end{figure}

Outside of the core, the FDM and N-body (CDM) profiles deviate by at most 50\% while the overall density varies by multiple orders of magnitude. To highlight the differences, we plot the residuals between the CDM and FDM halo profiles in \cref{fig:plot2b}.
The deviations are not correlated among different halos and may be caused by nonlinear amplification of numerical noise.

\begin{figure}
\includegraphics[width=\columnwidth]{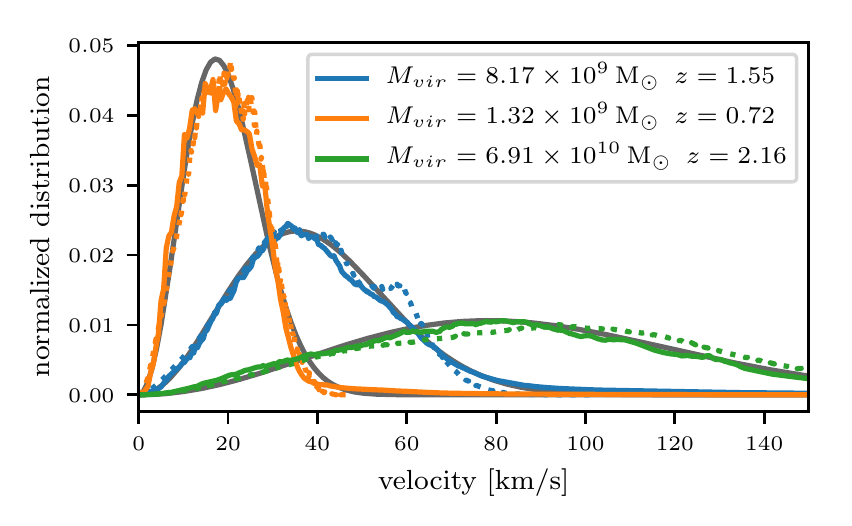}
\caption{Velocity distribution of plane waves in the Schr\"odinger field inside the virial radius (solid lines) and of particles in the same region (dotted lines). The grey solid lines show fitted Maxwellian distributions.} 
\label{fig:plot7}
\end{figure}

In order to compare FDM and CDM halos in momentum space, we consider the Wigner quasi-probability distribution
\begin{align}
f_\mathrm{W}(\mathbf{x},\mathbf{p}) = \int \frac{\text{d}^3 y}{(\pi\hbar)^3} \exp\left[ 2\frac{i}{\hbar} \mathbf{p}\cdot\mathbf{y}\right] \Psi(\mathbf{x} - \mathbf{y})\Psi^*(\mathbf{x}+\mathbf{y}) \,,
\end{align}
which matches the 6-dimensional phase space distribution function given by the Vlasov-Poisson equations, $f_\mathrm{CDM}$, when both are coarse-grained with a Gaussian filter obeying $\sigma_x\sigma_p \ge \hbar/2$ \cite{Widrow1993,Uhlemann2014}. The momentum distribution is thus obtained from the Fourier transform of $\Psi$,
\begin{align}
{f}_\mathrm{W}(\mathbf{p}) &= \frac{1}{N} \int \text{d}^3 x f_\mathrm{W}(\mathbf{x},\mathbf{p}) \cr 
&= \frac{1}{N} \left|\int \text{d}^3x \exp\left[-\frac{i}{\hbar} \mathbf{p} \cdot \mathbf{x}\right] \Psi(\mathbf{x})\right|^2 \,,
\end{align}
with $\mathbf{p} = m\mathbf{v}$ and a normalization factor $N$. In \cref{fig:plot7}, $f_\mathrm{W}(v)$ is compared to the velocity distribution $f(v)$ from the corresponding CDM simulations. As predicted by the Vlasov-Schr\"odinger correspondence, the normalized distribution of Fourier amplitudes matches very well the velocity distribution of particles in the N-body runs inside of the virial radius. Since the velocity distribution of virialised CDM halos is in rough approximation given by a Maxwellian distribution,
\begin{align}
\label{eq:maxwellian}
f(v) dv = \frac{4}{\pi} \left(\frac{3}{2}\right)^{3/2} \frac{v^2}{v_\mathrm{rms}^3} \exp\left(-\frac{3}{2} \frac{v^2}{v_\mathrm{rms}^2}\right) dv~,
\end{align}
where $v_\mathrm{rms}$ is the root-mean-square velocity \cite{Choi2014}, we also show Maxwellian distributions fitted to the Schr\"odinger results in \cref{fig:plot7}.

\begin{figure}
\includegraphics[width=\columnwidth]{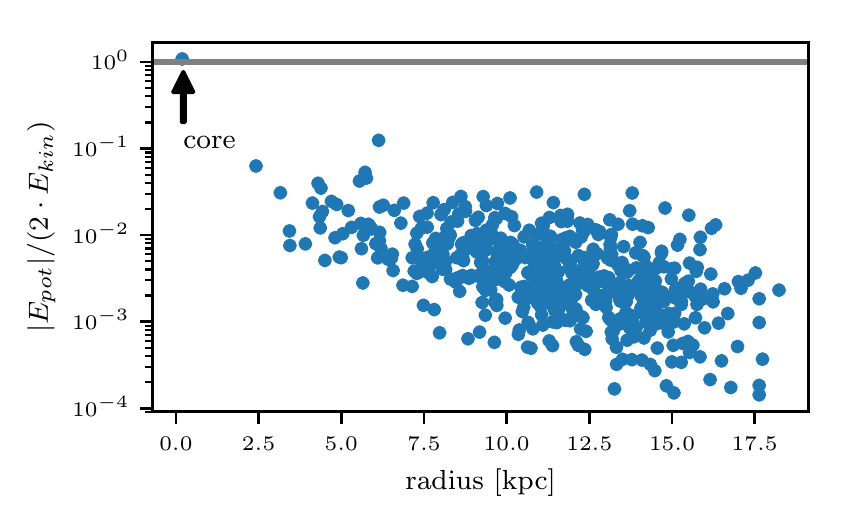}
\caption{Virial parameter of local maxima (granules) at various distances to the center of the halo.
} 
\label{fig:plot3}
\end{figure}

A powerful illustration of the difference between the core and the granular density fluctuations that make up the outer (incoherent) halo can be obtained by comparing their virial parameters (\cref{fig:plot3}). 
They are computed by taking spherical regions around local maxima in the density field and calculating the total kinetic and potential energy in these spheres. The radii of the spheres are given by the radius at which the angular averaged density drops to half its central value. The kinetic energy is computed by subtracting the center of mass velocity from the phase gradient and integrating
\begin{align}
E_\mathrm{kin} = \int \frac{\hbar^2}{2 m^2} |\nabla \Psi|^2 d^3 \mathbf{x}
\end{align}
over the volume of the sphere. The potential energy is approximated by the potential energy of a uniform sphere,
\begin{align}
E_\mathrm{pot} = - \frac{3}{5} \frac{G M^2(<r_{1/2})}{r_{1/2}}~.
\end{align}
This approximation is reasonable since the density profiles around local maxima are typically flattened within $r_{1/2}$. As can be seen in \cref{fig:plot3}, the core of the halo is the only local maximum that is in itself gravitationally bound and close to virialized. It is therefore a stable object whereas the granules have a finite lifetime of order $\tau = \hbar/m v_\mathrm{vir}^2$, as confirmed by the temporal correlation functions inside and outside of the core discussed below (cf. \cref{fig:plot8}).

\subsection{Time evolution}

\begin{figure}
\includegraphics[width=\columnwidth]{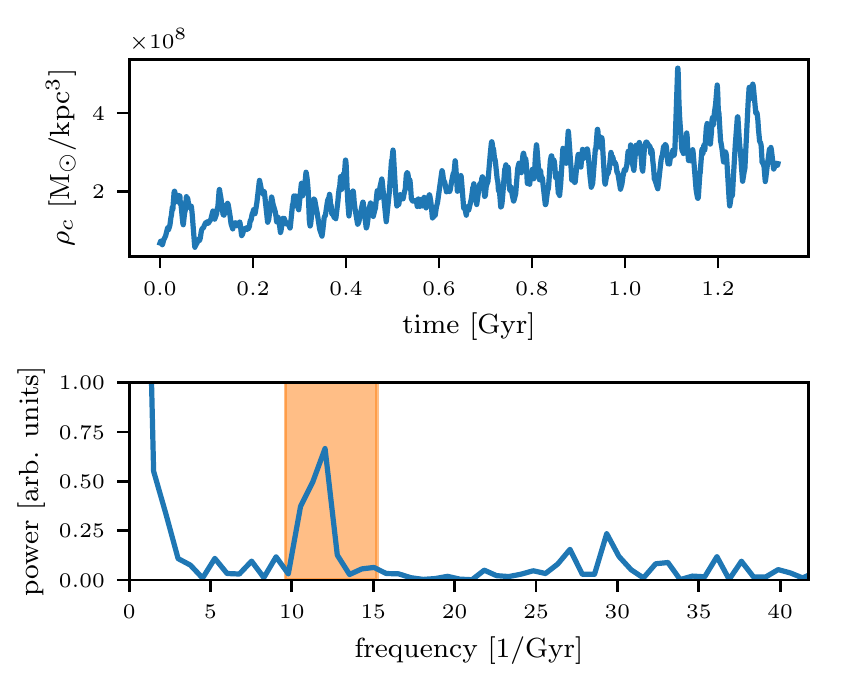}
\caption{Top: Maximum comoving density of a halo over time. Bottom: Fourier transform of the same data. The boundaries of the shaded region are the expected quasi-normal periods given the minimum and maximum central density in the time series above. } 
\label{fig:plot5}
\end{figure}

Schive \textit{et al.} \cite{Schive2014a} found the following relation between the core mass $M_\mathrm{c}$ and the total halo mass $M_\mathrm{h}$ (their Eq. 6):
\begin{align}
\label{eq:McMh}
M_\mathrm{c} = \frac{1}{4} a_\mathrm{vir}^{-1/2} \left( \frac{\zeta(a_\mathrm{vir})}{\zeta(0)} \right)^{1/6} \left( \frac{M_\mathrm{h}(a_\mathrm{vir})}{M_{0}} \right)^{1/3} M_{0}\,,
\end{align}
with $\zeta(a)$ from \cref{eq:zeta} and $M_0 \sim 4.4 \times 10^7 m_{22}^{-3/2} M_\odot$. For comparison with \cite{Schive2014a}, we also define $M_\mathrm{c}$ as the mass enclosed by the radius $x_\mathrm{c}$ where the peak density drops by a factor of $1/2$ and the density is assumed to follow a ground-state soliton profile (their Eq. 3). As discussed below, this is only approximately true owing to the strong oscillations of the core (see \cref{fig:plot5}). 

\begin{figure}
\includegraphics[width=\columnwidth]{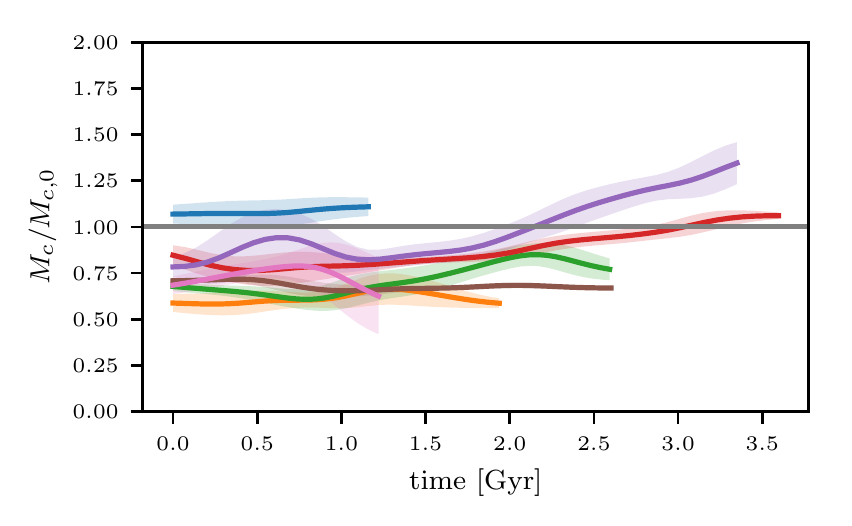}
\caption{Core masses from simulated halos normalized by  \cref{eq:McMh} at formation time as a function of halo age. The data points are smoothed in time with a Gaussian filter with $\sigma = 0.3$ Gyr. The shaded area represents the local standard deviation associated with the smoothing process.}
\label{fig:plot4}
\end{figure}

Note that we define $M_\mathrm{c}$ using fixed values for $a$, evaluated roughly at the time of halo virialization, instead of using the time-dependent scale factor and halo mass as done in \cite{Schive2014a}. This is motivated by our current understanding of the dynamics of core formation which determines $M_\mathrm{c}$ (cf. \cref{sec:discussion}). \Cref{fig:plot4} shows the evolution of core masses from our sample of halos as a function of time, normalized to \cref{eq:McMh}. Using the time-dependent values for $a$, $\zeta$ and $M_\mathrm{h}$ for the normalization produces differences that are small and unrelated to the halo mass. The time of virialization is determined by the requirement that the measured virial mass has settled to a slowly varying value. Spurious fluctuations of $M_\mathrm{c}$ resulting from oscillations of the peak density on much smaller timescales are smoothed by taking a moving average.
As can be seen, there is a small tendency towards smaller core masses than predicted by \cref{eq:McMh}.
No systematic growth of $M_\mathrm{c}$ by relaxation is observed for the majority of cores. Whether or not the mass increase of two of our simulated cores is related to ongoing condensation cannot be unambiguously answered at this point.

Analysis of the core density with much finer temporal resolution reveals oscillations with amplitudes of more than a factor  of two and a standard deviation of 33\% the mean density close to the dynamical timescale of the core (\cref{fig:plot5}). The frequency spectrum exhibits a peak at the quasi-normal frequency \cite{Guzman2004}
\begin{align}
 f = 10.94 \left(\frac{\rho_c}{10^9 \,\text{M}_\odot \text{kpc}^{-3}}\right)^{1/2} \text{Gyr}^{-1}\,,
\end{align}
with the central soliton density $\rho_c$. 
We thus find that cores form in a state with strong quasi-normal excitations, failing to relax to the ground state by gravitational cooling on evolutionary timescales. This result may open up new directions for observational probes of FDM cores.

\subsection{Correlation functions}

\begin{figure}
\includegraphics[width=\columnwidth]{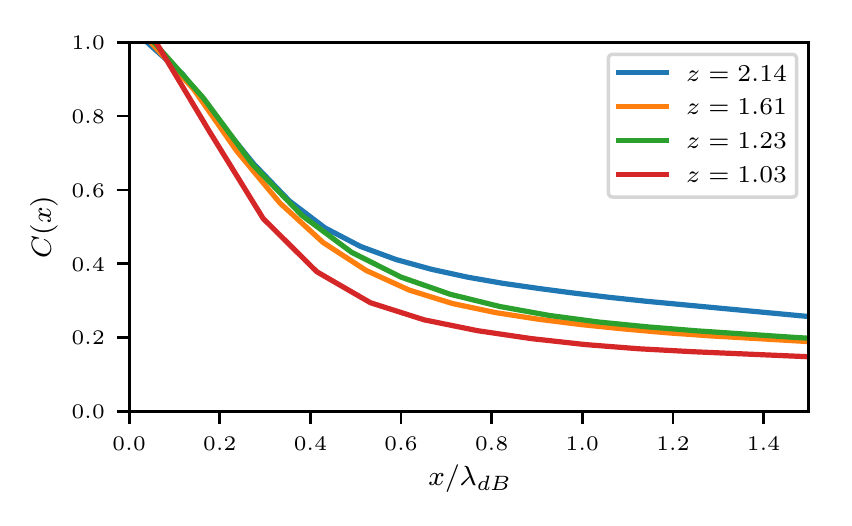}
\includegraphics[width=\columnwidth]{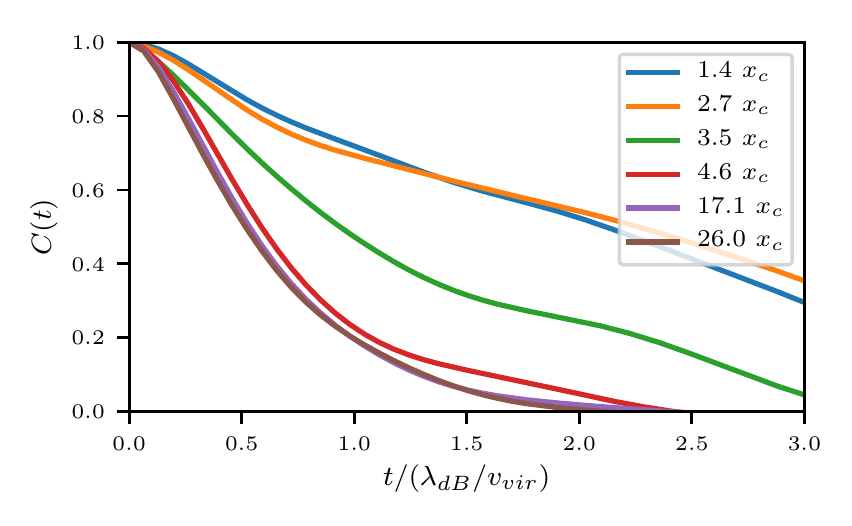}
\caption{Autocorrelation function of the density field in space (\textit{top}) and time (\textit{bottom}) inside the virial radius of a halo. The time correlation function is averaged over locations inside a given radial bin at some factor of the core radius $x_\mathrm{c}$. The coordinate system is comoving with the center of mass of the halo. Space and time units are normalized to the de Broglie wavelength corresponding to the virial velocity of the halo.} 
\label{fig:plot8}
\end{figure}

The spatial correlation function normalized to the virial de Broglie scale of the halo, $\lambda_\mathrm{dB} = \hbar/m v_\mathrm{vir}$,
\begin{align}
    C(x) = \frac{\langle \delta(\mathbf{x_1}) \delta(\mathbf{x_2}) \rangle_\mathbf{x}}{\langle \delta^2\rangle_\mathbf{x}}\,,
\end{align}
with $x = |\mathbf{x_1} - \mathbf{x_2}|$ and $\delta(\mathbf{x}) = \rho(\mathbf{x}) - \langle \rho \rangle_\mathbf{x}$ for a fixed halo at different redshifts can be seen in the top panel of \cref{fig:plot8}. As expected, the correlation length is of order $\lambda_\mathrm{dB}$ across a large range of redshifts. 

The temporal correlation function in the bottom panel of \cref{fig:plot8} is defined as
\begin{align}
    C(t,r) = \frac{\langle\langle \delta(t_1,\mathbf{x}) \delta(t_2,\mathbf{x}) \rangle_t\rangle_\mathbf{x}}{\langle\langle \delta^2(\mathbf{x})\rangle_t\rangle_\mathbf{x}}\,,
\end{align}
with $t = |{t_1} - {t_2}|$, $\delta(t, \mathbf{x}) = \rho(t, \mathbf{x}) - \langle \rho(\mathbf{x}) \rangle_\mathbf{t}$, $\langle\rangle_t$ denoting the temporal average, and $\langle\rangle_\mathbf{x}$ the spatial average within a radial bin with distance $r$ to the center. $\mathbf{x}$ is comoving with the halo's center of mass. The temporal correlation function confirms the enhanced coherence of the core with respect to the incoherent halo. Again, the curves are normalized to the coherence timescales expected for halo virialization, $\tau_\mathrm{c} = \hbar/m v_\mathrm{vir}^2$. The transition between the regions of high and low temporal coherence occurs at around $r = 3.5 x_\mathrm{c}$, which was previously found to be the radius where the solitonic radial profile turns into an NFW-like radial profile \cite{Mocz2017}.

\section{Discussion}
\label{sec:discussion}

\subsection{Core formation and evolution}
\label{sec:core-formation}

The results of our simulations present a consistent picture of the structure and dynamics of FDM halos and their cores. When coarse-grained on scales greater than the coherence length, the classical phase-space distribution of fields governed by the SP equations approaches that of a collisionless self-gravitating gas of particles. As is well-known from CDM simulations and shown above for FDM, the virialized velocity distribution is in good approximation given by a Maxwellian distribution, \cref{eq:maxwellian}, peaking at $v_{\rm rms} \sim v_{\rm vir}$ with the virial velocity $v_{\rm vir} \sim (M_{\rm h} / r_{\rm vir})^{1/2} \sim a^{-1/2} M_{\rm h}^{1/3}$. If a mode with the wavenumber $k$ locally exceeds the amplitude $\Psi_{\rm sol} (k) \sim (\hbar/m) k^2$, a self-bound, coherent solitonic state can form. This is most likely to occur at the center of the halo where the density is maximal, and the probability peaks at $k_{\rm vir} \sim m v_{\rm vir}/\hbar$ since the spectrum of $\Psi$ is given by \cref{eq:maxwellian}. If multiple solitons form initially, they rapidly merge into a single remnant with mass of the order of the most massive one \cite{Schwabe2016}. The outcome is a solitonic core with radius $R_{\rm c} \sim k_{\rm vir} ^{-1}$ exhibiting strong quasi-normal oscillations, embedded in a halo of incoherent field modes that follows a CDM-like density profile for radii $r \gg R_{\rm c}$.  The core mass is given by $M_{\rm c} \sim \Psi_{\rm sol}^2 k_{\rm vir}^{-3} \sim (\hbar/m)^2 k_{\rm vir}$ which explains the halo-core mass relation $M_{\rm c} \propto a^{-1/2} (\hbar/m) M_{\rm h}^{1/3}$. Accounting for the redshift dependence of overdensities and geometrical factors yields \cref{eq:McMh}, in agreement with \cite{Schive2014a}. 

In a very interesting recent paper by Levkov \textit{et al.} \cite{Levkov2018}, the formation of Bose stars is described in terms of classical wave condensation with gravitational interactions. While their model is consistent with our observation (first made by Schive \textit{et al.} \cite{Schive2014}) that solitonic cores form during the gravitational timescale of the halo, it is not clear whether it can explain the core-halo mass relation \cref{eq:McMh} and the strongly excited state of the core (\cref{fig:plot5}). Moreover, the homogeneous kinetic regime considered in \cite{Levkov2018} may not apply in our situation, see below. Comparing the time dependence of $M_\mathrm{c}$ predicted by this approach with our simulations is an interesting direction for future work.

Our results regarding the evolution of $M_\mathrm{c}$ on cosmological timescales are still inconclusive. Two out of seven simulated cores show continuing mass growth after formation (\cref{fig:plot4}). We verified that it is unrelated to the time dependent quantities in \cref{eq:McMh} ($a$, $\zeta$, and $M_\mathrm{h}$), confirming that $M_\mathrm{c}$ is determined by the formation process itself. A growing core mass could be understood in terms of condensation \cite{Levkov2018} or, more phenomenologically, by two-body relaxation of quasiparticles produced by the wave-like granularity of the incoherent halo \cite{Hui2017}. In any case, the simulations show only weak time dependence of $M_\mathrm{c}$ after formation, consistent with the observation that the coherence time of granular wavepatterns in the incoherent halo, $\tau_\mathrm{c} \sim \hbar m^{-1} v_\mathrm{vir}^{-2}$ (cf. \cref{fig:plot8}), is short compared to the halo dynamical timescale. This suggests that granular quasiparticles decay too quickly to experience any significant two-body momentum exchange. Note, however, that we have not yet explicitly explored most of the phenomena related to enhanced relaxation suggested in \cite{Hui2017}; this is the subject of ongoing work. 

The equivalence of the density and velocity distribution of the incoherent halo with the behavior of CDM halos comes as no surprise. As first pointed out in the context of cosmological simulations by Widrow and Kaiser \cite{Widrow1993} and explored as an alternative to N-body methods by Uhlemann \textit{et al.} \cite{Uhlemann2014} (see also \cite{Kopp2017,Mocz2018}), both are governed by the VP equations on scales larger than the coherence length. Specifically, the Wigner transform of a field described by the SP equations, coarse-grained by a Gaussian filter of width $\sigma_x$ and $\sigma_p$ with $\sigma_x\sigma_p \ge \hbar/2$ obeys the equally coarse-grained VP equations to first order in $\sigma_x^2$ and $\sigma_p^2$. Therefore, any statistical quantity averaged on scales greater than $\sigma_x$ and $\sigma_p$ can be expected to be indistinguishable between CDM and FDM. By directly comparing FDM and CDM simulations, our results provide direct evidence for this fact in the context of cosmological structure formation. On the other hand, we do not propose to invoke the SP-VP equivalence on scales of the coherent soliton to predict $M_\mathrm{c}$ (cf. \cite{Mocz2018}) because local nonlinear effects are not adequately captured by the coarse-grained description.

\subsection{Relation to wave turbulence and incoherent solitons}
\label{sec:wave-turb-incoh}

Systems described by classical wave dynamics have been investigated successfully using a kinetic formulation based on the concepts of wave turbulence \cite{Zakharov1992,Nazarenko2011}. One particular example is the field of statistical nonlinear optics, see \cite{Garnier2012,Picozzi2014} for an overview. Different types of kinetic equations apply depending on the degree of spatial or temporal homogeneity of the wave statistics and the level of nonlocality of the nonlinear interaction. 

For instance, a system with local nonlinearity that is statistically homogeneous in space is governed by the wave turbulence kinetic equation describing the statistical behavior of random weakly nonlinear waves. The structure of this kinetic equation is analogous to the Boltzmann kinetic equation for a dilute gas. In the case of wave turbulence, four-wave interactions give rise to an irreversible evolution toward a Rayleigh-Jeans distribution and classical condensation of a long-wavelength coherent mode, in complete analogy to Bose-Einstein condensation \cite{Connaughton2005}. 

On the other hand, the gravitational interaction in the SP equations is characterized by long-range nonlocality (in non-comoving coordinates):
\begin{align}
\label{eq:SP}
 i \hbar  \frac{\partial \Psi}{\partial t} = \left[ -\frac{\hbar^2}{2 m} \nabla^2 - G m \int \frac{ \vert\Psi(\mathbf{x'})\vert^2}{\vert\mathbf{x}-\mathbf{x'}\vert} \, d^3 x' \right]\, \Psi\,\,.
\end{align}
As shown in \cite{Picozzi2011} in the context of nonlinear optics, Schrödinger systems with long-range nonlocal nonlinearities are subject to the modulational instability (translating into gravitational instability in our context) and form \emph{incoherent solitons} trapped by a self-consistent potential corresponding to (incoherent) dark matter halos in FDM cosmology. Moreover, the nonlocality strongly suppresses thermalization and condensation. The growth of modulational/gravitational instabilities renders the wave statistics inhomogeneous which may quench the condensation described in \cite{Levkov2018} on the scales of FDM cores. 

Again, this behavior is related to the analogy with collisionless gases with long-range interactions described by the Vlasov equation. \cite{Garnier2012} show that the Wigner transform of the autocorrelation function of $\Psi$ obeys a Vlasov equation to first order in a multiscale expansion, similar to the coarse-graining approach used by \cite{Widrow1993,Uhlemann2014}. The self-consistent potential for a gravity-like nonlinearity is given by 
\begin{equation}
V = G \int \frac{ \langle \vert\Psi(\mathbf{x'})\vert^2 \rangle}{\vert\mathbf{x}-\mathbf{x'}\vert} \, d^3 x' \,\,,
\end{equation}
where the squared field amplitude is averaged over an appropriate intermediate-scale spatial domain. It is also shown that in the highly nonlocal limit,  $V$ can be treated as a fixed potential, and \cref{eq:SP} reduces to a linear Schrödinger equation describing waves that propagate in this background. Crucially, thermalization toward thermodynamic equilibrium is absent for long-range nonlocal interactions at this level of approximation,
a feature which is consistent with the formal reversibility of the Vlasov equation.

\section{Conclusions}
\label{sec:conclusions}

Using zoom-in simulations with a hybrid N-body and finite difference method to solve the coupled Schrödinger-Poisson (SP) equations, we studied the formation and time evolution of halos composed of ultralight bosonic dark matter (FDM) from cosmological initial conditions. Our sample contains seven halos with masses ranging from $8 \times 10^8$ M$_\odot$ and $7 \times 10^{10}$ M$_\odot$. We confirm the general structure of FDM halos of coherent solitonic cores embedded in incoherent halos  reported by \cite{Schive2014,Schive2014a}, as well as their core-halo mass relation \cref{eq:McMh} evaluated at the time of core formation with a small tendency towards lower core masses than predicted.

Considering the core mass on cosmological timescales, we found no conclusive indications of mass growth by condensation. The maximum overall mass increase in our sample of cores was 70 \% over a period of $\sim 3$ Gyr. The core mass does not obviously follow the time-dependent quantities in the core-halo mass relation. It appears to be governed chiefly by the coherence length of the FDM field at the time of virialization of the host halo which determines the core radius and thus its mass. 

The simulated cores form in a highly excited state with strong quasi-normal oscillations that do not decay on the halo evolutionary timescale. These oscillations might give rise to new observational probes for the existence of FDM cores, providing an independent determination of the core mass if the oscillation frequency can be measured.  

The radial density profile of the incoherent halo surrounding the core shows only small and local deviations from comparison N-body runs for corresponding CDM halos. This, as well as the approximate Maxwellian distribution of the velocity spectrum, is consistent with the equivalence of the SP and Vlasov-Poisson (VP) equations on scales far greater than the coherence length. We were unable to detect any signatures of enhanced gravitational relaxation as proposed by \cite{Hui2017} but note that further work is required for a definitive answer. 

We only considered halos without major mergers in this work. The core-halo mass relation for merger-dominated systems was predicted by \cite{Du2017b} based on the mass change in individual core mergers \cite{Schwabe2016}. Halos with higher masses and merger activity require higher resolution simulations. We hope to return to this question after further optimizations.

One of the most interesting open questions is the behavior of FDM cores and halos in the presence of baryons. This will be addressed in forthcoming work.

\acknowledgements
We thank Katy Clough, Xiaolong Du, Shaun Hotchkiss, Erik Lentz, Doddy Marsh, and Javier Redondo for helpful discussions and comments. JCN would additionally like to thank Antonio Picozzi and acknowledges the Aspen Center for Physics, supported by National Science Foundation grant PHY-1607611, where parts of this work were done.
Computations described in this work were performed using the
publicly-available \textsc{Enzo} code (http://enzo-project.org).
The simulations were performed with resources provided by the North-German Supercomputing Alliance (HLRN).
We acknowledge the \textsc{yt} toolkit \cite{Turk2011} that was used for our analysis of numerical data.

\end{document}